\documentclass[nofootinbib, twocolumn, preprintnumbers]{revtex4-1}
\pdfoutput=1

\usepackage{amsmath, amssymb, graphicx, booktabs, bm, psfrag, color}
\usepackage{array}

\begin{document}

\title{\textbf{Explaining the 750 GeV diphoton excess with a colored scalar\\
    charged under a new confining gauge interaction}}

\preprint{April, 2016}

\author{Robert Foot}
\email{rfoot@unimelb.edu.au}
\affiliation{ARC Centre of Excellence for Particle Physics at the
  Terascale,\\School of Physics, The University of Melbourne, Victoria 3010, Australia}
\author{John Gargalionis}
\email{garj@student.unimelb.edu.au}
\affiliation{ARC Centre of Excellence for Particle Physics at the
  Terascale,\\School of Physics, The University of Melbourne, Victoria 3010, Australia}

\begin{abstract}
  \noindent We consider a charged scalar particle $\chi$ of mass around 375
  GeV charged under both $\mathrm{SU}(3)_{\textsc{c}}$ and a new confining
  non-abelian gauge interaction. After
  pair production, these interactions confine the exotic scalar into
  non-relativistic bound states whose decays into photons can explain the 750 GeV diphoton excess observed at the LHC.
  Taking the new confining group to be $\mathrm{SU}(2)$, we find
  $\chi$ must carry an electric charge of $Q \sim [\frac{1}{2}, 1]$ to fit the data.
  Interestingly, we find that pair production of the scalars and the subsequent
  formation of the bound state dominates over direct bound state resonance
  production. This explanation is quite weakly constrained by current searches
  and data from the forthcoming run at the LHC will be able to probe our
  scenario more fully. In particular
  dijet, mono-jet, di-Higgs and $\text{jet} + \text{photon}$ searches may be the most promising
  discovery channels. 
\end{abstract}

\maketitle

\section*{Introduction}

An excess of events containing two photons with invariant
mass near 750 GeV has been observed in 13 TeV proton--proton collisions by the
\textsc{atlas} and \textsc{cms} collaborations~\cite{atlas1, cms1}. The cross
section $\sigma(pp \rightarrow \gamma \gamma)$ is estimated to be
\begin{align}
  \begin{split}
  \sigma (p p \rightarrow \gamma \gamma) &=
  \begin{cases}
    (10 \pm 3) \text{ fb} & \textsc{atlas} \\
    (6 \pm 3) \phantom{1}\text{ fb} & \textsc{cms}
  \end{cases} 
  \end{split}
\end{align}
and there is no evidence of any accompanying excess in the dilepton
channel~\cite{atlas2}. If we interpret this excess as the two photon decay of a
single new particle of mass $m$ then \textsc{atlas} data provide a hint of a
large width: $\Gamma/m \sim 0.06$, while \textsc{cms} data prefer a narrow
width. Naturally, further data collected at the \textsc{lhc} should provide a
clearer picture as to the nature of this excess.
 
There has been vast interest in the possibility that the diphoton excess results
from physics beyond the standard model (\textsc{sm}). Most discussion has
focused on models where the excess is due to a new scalar particle which
subsequently decays into two photons e.g. \cite{np1} (for a recent discussion
see also~\cite{strumia}). The possibility that the new scalar particle is a
bound state of exotic charged fermions has also been considered, e.g.
\cite{b1,b2,b3,b4,b5}. Here we consider the case that the 750 GeV state
is a non-relativistic bound state constituted by an exotic \textit{scalar} particle $\chi$ and
its antiparticle, charged under $\mathrm{SU}(3)_{\textsc{c}}$ as well as a new
unbroken non-abelian gauge interaction. 
Having $\chi$ be a scalar rather than a fermion is not merely a matter of taste: In such a framework
a fermionic $\chi$ would lead to the formation of bound states which (typically) decay to 
dileptons more often than to photons; a situation which is not favoured by the data.

The bound state, which we denote $\Pi$,
can be produced through gluon--gluon fusion directly (i.e. at threshold
$\sqrt{s_{gg}} \simeq M_\Pi$) or indirectly via $gg \rightarrow \chi^\dagger
\chi \rightarrow \Pi + \textit{soft quanta}$ (i.e. above $\Pi$ threshold:
$\sqrt{s_{gg}} > M_\Pi$). The indirect production mechanism can dominate the
production of the bound state, which is an interesting feature of this kind of
theory.

\section*{The model}

We take the new confining unbroken gauge interaction to be $\mathrm{SU}(N)$, and
assume that, like $\mathrm{SU}(3)_{\textsc{c}}$, it is asymptotically free and
confining at low energies. However, the new $\mathrm{SU}(N)$ dynamics is
qualitatively different from \textsc{qcd} as all the matter particles (assumed
to be in the fundamental representation of $\mathrm{SU}(N)$) are taken to be
much heavier than the confinement scale, $\Lambda_{\textsc{n}}$. In fact we here
consider only one such matter particle, $\chi$, so that $M_\chi \gg
\Lambda_{\textsc{n}}$ is assumed. In this circumstance a $\chi^\dagger \chi$
pair produced at the \textsc{lhc} above the threshold $2M_\chi$ but below
$4M_\chi$ cannot fragment into two jets. The $\mathrm{SU}(N)$ string which
connects them cannot break as there are no light $\mathrm{SU}(N)$-charged states
available. This is in contrast to heavy quark production in \textsc{qcd} where
light quarks can be produced out of the vacuum enabling the color string to
break. The produced $\chi^\dagger\chi$ pair can be viewed as a highly excited
bound state, which de-excites by $\mathrm{SU}(N)$-ball and soft glueball/pion
emission~\cite{hall}.

With the new unbroken gauge interaction assumed to be $\mathrm{SU}(N)$ the gauge symmetry
of the \textsc{sm} is extended to
\begin{equation}
  \label{eq:gaugegroup}
\mathrm{SU(3)}_{\textsc{c}} \otimes \mathrm{SU}(2)_{\textsc{l}} \otimes \mathrm{U}(1)_{\textsc{y}} \otimes \mathrm{SU}(N).
\end{equation}
This kind of theory can arise naturally in models which feature large colour
groups~\cite{fh,f,tg} and in models with  leptonic colour~\cite{fl,fv,jfv}
but was also considered earlier by Okun~\cite{okun}. The notation
\textit{quirks} for heavy particles charged under an unbroken gauge symmetry
(where $M_\chi \gg \Lambda_{\textsc{n}}$) was introduced in~\cite{hall} where the relevant
phenomenology was examined in some detail in a particular model \footnote{
Some other aspects of such models have been discussed over the years, including the possibility that
the $\mathrm{SU}(N)$ confining scale is low ($\sim$ keV), a situation which leads to macroscopic strings \cite{kang}.}.
For convenience
we borrow their nomenclature and call the new quantum number \textit{hue} and
the massless gauge bosons \textit{huons} ($\mathcal{H}$).

The phenomenological signatures of the bound states (quirkonia) formed depend on
whether the quirk is a fermion or boson. Here we assume that the quirk $\chi$ is
a Lorentz scalar in light of previous work which indicated that bound states
formed from a fermionic $\chi$ state would be expected to be observed at the
\textsc{lhc} via decays of the spin 1 bound state into opposite-sign lepton
pairs ($\ell^+\ell^-$)~\cite{hall,jfv}. In fact, this appears to be a serious
difficulty in attempts to interpret the 750 GeV state as a bound state of
fermionic quirk particles (such as those of \cite{b2,b3,b4}). The detailed
consideration of a scalar $\chi$ appears to have been largely
overlooked\footnote{The idea has been briefly mentioned in recent
  literature~\cite{Agrawal:2015dbf, b4}.}, perhaps due to the paucity of known elementary
scalar particles. With the recent discovery of a Higgs-like scalar at 125
GeV~\cite{atlas3,cms3} it is perhaps worth examining signatures of scalar quirk
particles. In fact, we point out here that the two photon decay is the most
important experimental signature of bound states formed from electrically
charged scalar quirks. Furthermore this explanation is only weakly constrained
by current data and thus appears to be a simple and plausible option for the new
physics suggested by the observed diphoton excess.

\section*{Explaining the excess}
\begin{figure*}[t]
\begin{center}
\begin{tabular}{m{4.1cm} m{4.1cm} m{3.0cm} m{2.86cm} m{2.1cm}}
  \includegraphics[scale=0.92]{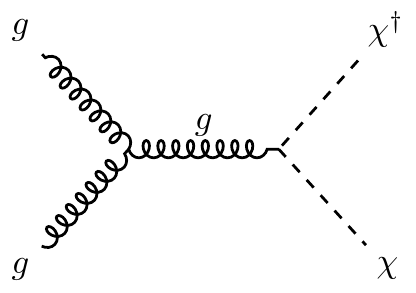}
  &
    \includegraphics[scale=0.92]{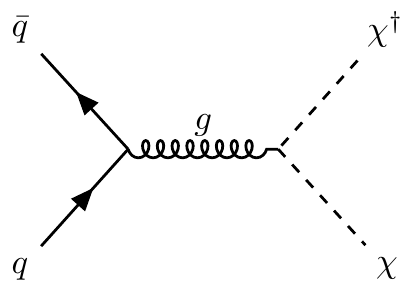}
  &
    \includegraphics[scale=0.92]{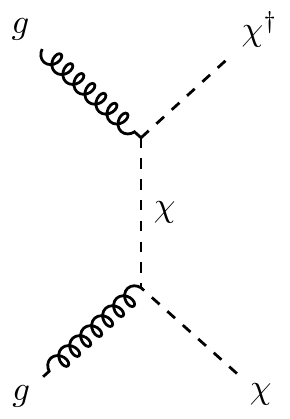}
  &
    \includegraphics[scale=0.92]{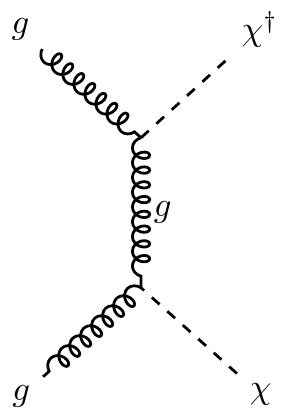}
  &
    \includegraphics[scale=0.92]{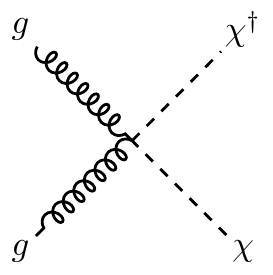}

\end{tabular}
\caption{Tree-level pair production mechanisms for the scalar quirk $\chi$.}
\label{fig:prod}
\end{center}
\end{figure*}

The scalar $\chi$ that we introduce transforms under the extended gauge group
(eq.~\ref{eq:gaugegroup}) as
\begin{equation}
  \chi \sim (\mathbf{3}, \mathbf{1}, Y; \mathbf{N}),
\end{equation}
where we use the normalization $Q=Y/2$. The possibility that $\chi$ also
transforms non-trivially under $\mathrm{SU}(2)_{\textsc{l}}$ is interesting,
however for the purposes of this letter we focus on the 
$\mathrm{SU}(2)_{\textsc{l}}$ singlet case for definiteness. Since two-photon
decays of non-relativistic quirkonium will be assumed to be responsible for
the diphoton excess observed at the LHC, the mass of $\chi$ will need to be
around 375 GeV.

We have assumed that $\chi$ is charged under $\mathrm{SU}(3)_{\textsc{c}}$ so
that it can be produced at tree-level through \textsc{qcd}-driven pair
production. We present the production mechanisms in fig.~\ref{fig:prod}. To
estimate the production cross section of the bound states, we first consider the
indirect production mechanism which we expect to be dominant. Here, a
$\chi^\dagger \chi$ pair is produced above threshold and de-excites emitting
soft glueballs/pions and hueballs: $gg \rightarrow \chi^\dagger \chi \rightarrow \Pi +
\textit{soft quanta}$. We first consider the case where the confinement scale of
the new $\mathrm{SU}(N)$ interaction is similar to that of \textsc{qcd}. What
happens in this case can be adapted from the discussion in \cite{hall}, where a
fermionic quirk charged under an unbroken $\mathrm{SU}(2)$ gauge interaction was
considered. As already briefly discussed in the introduction, the $\chi^\dagger
\chi$ pairs initially form a highly excited bound state, which subsequently
de-excites in two stages. The first stage is the non-perturbative regime where
the hue string is longer than $\Lambda_{\textsc{n}}^{-1}$. The second stage is
characterized by a string scale significantly less than
$\Lambda_{\textsc{n}}^{-1}$: the perturbative Coulomb region. Here the bound
state can be characterized by the quantum numbers $n$ and $l$. De-excitation
continues until quirkonium is in a lowly excited state with $l \leq 1$ and $n$.
Imagine first that de-excitation continued until the ground state ($n=1$, $l=0$)
is reached. Given we are considering $\chi$ to be a scalar, the 
quirkonium ground state, $\Pi$, will have spin 0, and is thus expected to decay into
\textsc{sm} gauge bosons and huons. The cross section $\sigma(pp \rightarrow \Pi
\rightarrow \gamma \gamma)$ in this case is then
\begin{equation}
\sigma (pp \rightarrow \gamma \gamma) \approx \sigma(pp \rightarrow \chi^\dagger
  \chi) \times \text{Br}(\Pi \rightarrow \gamma \gamma).
\end{equation}

Since production is governed by \textsc{qcd} interactions, we can use the values
of the pair production cross sections for stops/sbottoms in the limit of
decoupled squarks and gluinos~\cite{Borschensky:2014cia}. For a $\chi$ mass of
375 GeV
\begin{align}
  \begin{split} \label{eq:ind1}
  \sigma (p p \rightarrow \chi^\dagger \chi) &\approx
  \begin{cases}
    2.6 N \text{ pb} & \text{at 13 TeV} \\
    0.5 N \text{ pb} & \text{at \phantom{1}8 TeV}
  \end{cases}.
  \end{split}
\end{align}
The branching fraction is to leading order:
\begin{equation}
  \text{Br}(\Pi \rightarrow \gamma \gamma) \simeq \frac{3NQ^4 \alpha^2}{\frac{2}{3}N\alpha_{\textsc{s}}^2
    + \frac{3}{2}C_{\textsc{n}}\alpha_{\textsc{n}}^2 + 3NQ^4\alpha^2},
\label{5a}
\end{equation}
where $C_{\textsc{n}} \equiv (N^2 - 1)/(2N)$, $\alpha_{\textsc{n}}$ is the new
$\mathrm{SU}(N)$ interaction strength and we have neglected the small
contribution of $\Pi \rightarrow Z\gamma / ZZ$ to the total width. Eq.~\ref{5a}
also neglects the decay to Higgs particles: $\Pi \to hh$, which arises from the
Higgs potential portal term $\lambda_\chi \chi^\dagger \chi \phi^\dagger \phi$.
Theoretically this rate is unconstrained given the dependence on the unknown
parameter $\lambda_\chi$, but could potentially be important. However, limits
from resonant Higgs boson pair production derived from 13 TeV data: $\sigma (pp
\rightarrow X \rightarrow hh \rightarrow bbbb) \lesssim 50$ fb at $M_X \approx
750$ GeV \cite{atlashh,cmshh} imply that the Higgs decay channel must indeed be
subdominant (cf. $\Pi \rightarrow gg$, $\mathcal{H}\mathcal{H}$).

\begin{figure*}[t]
\begin{center}
  \includegraphics[scale=0.48]{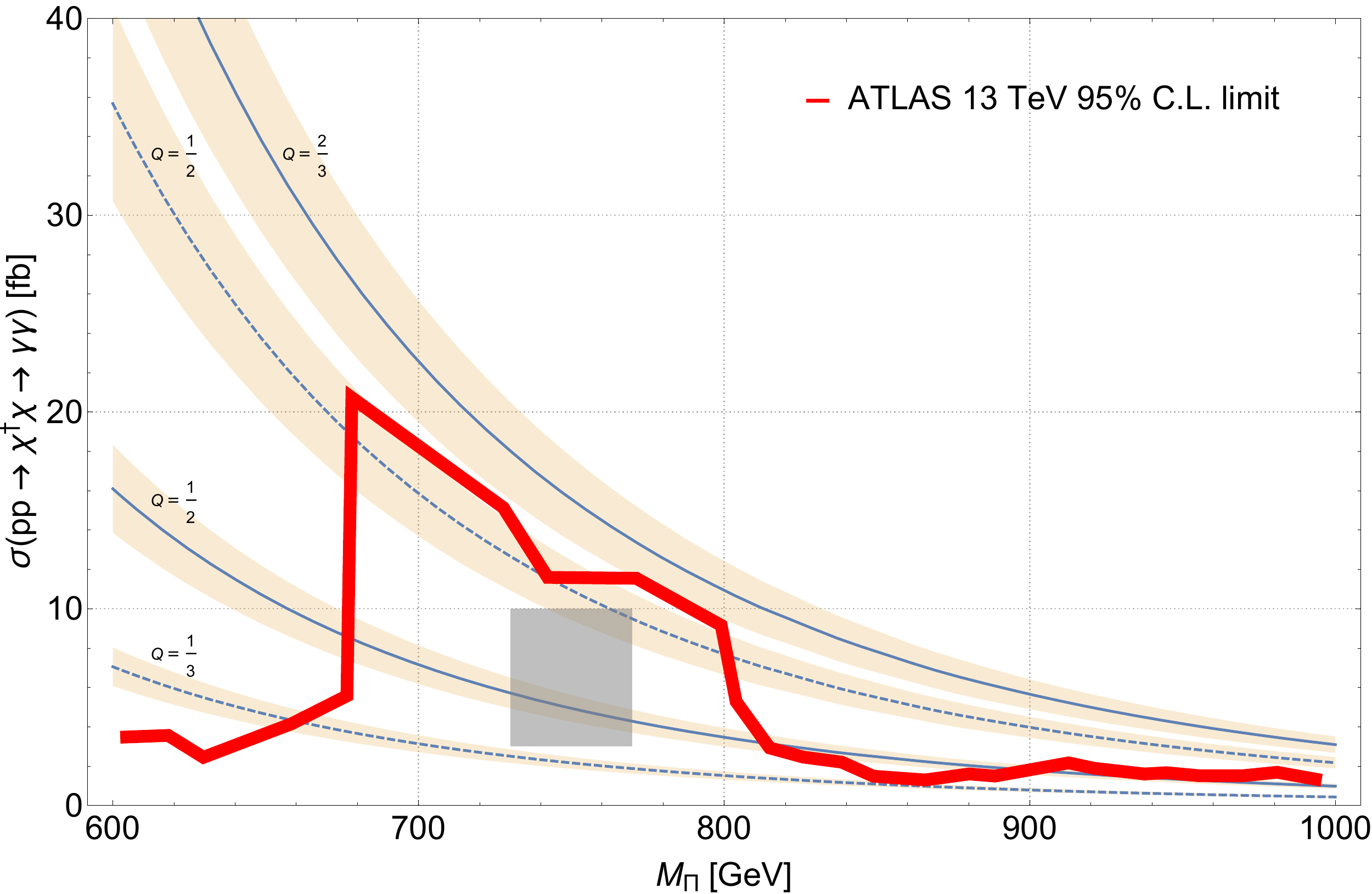}
  \caption{The cross section $\sigma(pp \rightarrow \Pi \rightarrow \gamma
    \gamma)$ at 13 TeV for a range of quirkonium masses $M_\Pi$ and charge
    assignments. Solid lines denote choices of $N=2$ and dashed lines choices of
    $N=5$. 
    The rectangle represents the $\sigma \in [3, 10]$ fb indicative region
    accommodated by the \textsc{atlas} and \textsc{cms} data. The solid red line
    is the \textsc{atlas} 13 TeV exclusion limit. Uncertainties reflect error
    associated with the parton distribution functions.}
\label{fig:plot}
\end{center}
\end{figure*}

The renormalized gauge coupling constants in eq.~\ref{5a} are evaluated at the
renormalization scale $\mu \sim M_\Pi/2$. Taking for instance the specific case
of $N=2$, $\alpha_{\textsc{n}} = \alpha_{\textsc{s}} \simeq 0.10$ (at $\mu \sim
M_\Pi/2$) gives
\begin{equation}
\sigma (pp \rightarrow \gamma \gamma) \approx 5 \left( \frac{Q}{1/2} \right)^4 \text{ fb}
\ \ \ \text{at 13 TeV}.
\end{equation}
At $\sqrt{s} = 8$ TeV the cross section is around five times smaller. We present the cross
section $\sigma(pp \rightarrow \Pi \rightarrow \gamma \gamma)$ for a range of
masses $M_\Pi$ and different combinations of $Q$ and $N$ in fig.~\ref{fig:plot}.
The parameter choice $\alpha_{\textsc{n}}=\alpha_{\textsc{s}}$ and
$\Lambda_{\textsc{n}}=\Lambda_{\textsc{qcd}}$ has been assumed. (The cross
section is not highly sensitive to $\Lambda_{\textsc{n}}$, $\alpha_{\textsc{n}}$
so long as we are in the perturbative regime: $\Lambda_{\textsc{n}}\lesssim
\Lambda_{\textsc{qcd}}$.) Evidently, for $N = 2$, a $\chi$ with electric charge
$Q \approx 1/2$ is produced at approximately the right rate to explain the
diphoton excess.  

In practice de-excitation of the produced quirkonium does not always continue
until the ground state is reached. In this case annihilations of excited states
can also contribute. However those with $l=0$ will decay in the same way as the
ground state. The only difference is that the excited states will have a
slightly larger mass (which we will estimate in a moment) due to the change in the
binding energy. This detail could be important as it can effectively enlarge the
observed width. Annihilation of excited states with non-zero orbital angular
momentum could in principle also be important, however these are suppressed as
the radial wavefunction vanishes at the origin: $R(0) = 0$ for $l \geq 1$. They
are expected to de-excite predominately to $l=0$ states rather than
annihilate~\cite{hall}. Nevertheless, for sufficiently large
$\alpha_{\textsc{n}}$ the $l=1$ annihilations: $\Pi \rightarrow \mu^+\mu^-$ and
$\Pi \rightarrow e^+e^-$ could potentially be observable.

The $l=0$ excited states can be characterized by the quantum number $n$ with
binding energies:
\begin{equation}
  \frac{E_n}{M_\Pi} = - \frac{1}{8n^2} \left[ \frac{4}{3} \bar{\alpha}_{\textsc{s}} + C_{\textsc{n}} \bar{\alpha}_{\textsc{n}} + Q^2 \bar{\alpha} \right]^2.
\label{eq:8y}
\end{equation}
The above formula was adapted from known results with quarkonium, e.g.
\cite{kats} (and of course also the hydrogen atom). The coupling constants
$\bar{\alpha}_{\textsc{s}}$, $\bar{\alpha}_{\textsc{n}}$ and $\bar{\alpha}$ are
evaluated at a renomalization scale corresponding to the mean distance between
the particles which is of order the Bohr radius: $a_0 =
4/[(4\bar{\alpha}_{\textsc{s}}/3 + C_{\textsc{n}}\bar{\alpha}_{\textsc{n}} + Q^2
\bar{\alpha})M_\Pi]$. The bound state, described by the radial quantum number
$n$ has mass given by $M_\Pi(n) = 2M_\chi + E_n$. Considering as an example
$N=2$ and $\bar{\alpha}_{\textsc{n}} = \bar{\alpha}_{\textsc{s}} = 0.15$,
$\bar{\alpha} = 1/137$ we find the mass difference between the $n=1$ and $n=2$
states to be $\Delta M = (E_1-E_2) \approx 0.01 M_\Pi$. Larger mass splitings
will be possible\footnote{Additional possibilities arize if $\chi$ transforms
  nontrivially under $\mathrm{SU}(2)_{\textsc{l}}$, i.e. forming a
  representation $\mathbf{N}_{\textsc{l}}$. The mass degeneracy of the multiplet
  will be broken at tree-level by Higgs potential terms along with electroweak
  radiative corrections. The net effect is that the predicted width of the $pp
  \rightarrow \gamma \gamma$ bump can be effectively larger as there are
  $\mathrm{N}_{\textsc{l}}$ distinct bound states, $\Pi^i$, (of differing
  masses) which can each contribute to the decay width. Although each state is
  expected to have a narrow width, when smeared by the detector resolution the
  effect can potentially be a broad feature.} if $\bar{\alpha}_{\textsc{n}} >
\bar{\alpha}_{\textsc{s}}$, although it has been shown in the context of
fermionic quirk models that the phenomenology is substantially altered in this
regime~\cite{b2}. In particular, the hueballs can become so heavy that the
decays of the bound state into hueballs is kinematically forbidden.

In the above calculation of the bound state production cross section, we
considered only the \emph{indirect} production following pair production of
$\chi^\dagger \chi$ above threshold. The bound state can also be produced
directly: $gg \rightarrow \Pi$, where $\sqrt{s_{gg}} \approx M_\Pi$. The cross
section of the ground state direct resonance production is
\begin{equation}
\sigma (pp \rightarrow \Pi)_{\textsc{dr}} \approx \frac{C_{gg} K_{gg} \Gamma (\Pi \rightarrow gg)}{sM_\Pi},
\end{equation}
where $C_{gg}$ is the appropriate parton luminosity coefficient and $K_{gg}$ is
the gluon \textsc{nlo} \textsc{qcd} K-factor. For $\sqrt{s} = 13$ TeV we take
$C_{gg} \approx 2137$~\cite{np1} and $K_{gg} = 1.6$~\cite{Harlander:2005rq}. The
partial width $\Gamma (\Pi \rightarrow gg)$ of the $n=1, \ l=0$ ground state is
given by
\begin{equation}
  \Gamma (\Pi \rightarrow gg) =  \frac{4}{3} M_\Pi N \alpha_{\textsc{s}}^2
  \frac{|R(0)|^2}{M_\Pi^3} ,
\end{equation}
where
the radial wavefunction at the origin for the ground state is:
\begin{equation}
\frac{|R(0)|^2}{M_\Pi^3} = \frac{1}{16} \left[ 
\frac{4}{3} \bar \alpha_s + C_{\textsc{n}} \bar{\alpha}_{\textsc{n}} + Q^2 \bar{\alpha}
\right]^3.
\label{eq:11y}
\end{equation}
Considering again the example of $N=2$ and $\bar{\alpha}_{\textsc{n}} =
\bar{\alpha}_{\textsc{s}} = 0.15$, $\bar{\alpha} = 1/137$ we find
\begin{equation}
  \sigma(pp \rightarrow \Pi)_{\textsc{dr}} \approx 0.40 \text{ pb} \ \ \ \text{at 13 TeV}.
\end{equation}
Evidently, the direct resonance production cross section is indeed expected to be subdominant, around 8\% that of the 
indirect production cross section
(eq.~\ref{eq:ind1}) 
\footnote{If
  $\bar{\alpha}_{\textsc{n}}$ is sufficiently large, one can potentially have
  direct resonance production comparable or even dominating indirect production
  (such a scenario has been contemplated recently in \cite{b3,b4}). Naturally at
  such large $\bar{\alpha}_{\textsc{n}}$ the perturbative calculations become
  unreliable, and one would have to resort to non-perturbative techniques such as lattice computations.}.

We now comment on the regime where $\Lambda_{\textsc{n}}$ is smaller than
$\Lambda_{\textsc{qcd}}$. In fact, if the $\mathrm{SU}(N)$ confining scale is
only a little smaller than $\Lambda_{\textsc{qcd}}$ then a light quark pair can
form out of the vacuum, leading to a bound state of two \textsc{qcd} color
singlet states: $\chi \bar{q}$ and $\chi^\dagger q$. These color singlet states
would themselves be bound together by $\mathrm{SU}(N)$ gauge interactions to
form the $\mathrm{SU}(N)$ singlet bound state. Since only $\mathrm{SU}(N)$
interactions bind the two composite states ($\chi \bar{q}$ and $\chi^\dagger
q$), it follows that $\frac{4}{3} \bar \alpha_s + C_{\textsc{n}}
\bar{\alpha}_{\textsc{n}} + Q^2 \bar{\alpha} \to C_{\textsc{n}} \bar{\alpha}_{\textsc{n}} + (Q - Q_q)^2 \bar{\alpha}$ in
eqs.~\ref{eq:8y} and \ref{eq:11y}. Therefore if the confinement scale of
$\mathrm{SU}(N)$ is smaller than that of \textsc{qcd} then the direct production
rate becomes completely negligible relative to the indirect production
mechanism. The rate of $\Pi$ production is the same as that found earlier in
eq.~\ref{eq:ind1}, but the branching ratio to two photons is modified:
\begin{equation}
\text{Br}(\Pi \rightarrow \gamma \gamma) \simeq \frac{3NQ^4 \alpha^2}{\frac{7}{3}N\alpha_{\textsc{s}}^2
+ \frac{3}{2} C_{\textsc{n}} \alpha_{\textsc{n}}^2 + 3NQ^4\alpha^2},
\end{equation}
where, as before, we have neglected the small contribution of $\Pi \rightarrow
Z\gamma / ZZ$ to the total width, and also the contribution from $\Pi \rightarrow hh$. 
In this regime somewhat larger values of $Q$ can be accommodated, such as $Q = 5/6$ for $N=2$ \footnote{
Although it is perhaps too early to speculate on the possible role of $\chi$ in a more elaborate framework,
we nevertheless remark here that particles fitting its description are required for spontaneous symmetry breaking of
extended Pati-Salam type unified theories \cite{eps}.}.

Notice that in the $\Lambda_{\textsc{n}} < \Lambda_{\textsc{qcd}}$ regime the
size of the mass splittings between the excited states becomes small as
$\frac{4}{3} \bar \alpha_s + C_{\textsc{n}} \bar{\alpha}_{\textsc{n}} + Q^2 \bar{\alpha} \to
C_{\textsc{n}} \bar{\alpha}_{\textsc{n}} + (Q - Q_q)^2 \bar{\alpha} $ in eq.~\ref{eq:8y}. We therefore
expect no effective width enhancement due to the excited state decays at the LHC
in the small $\Lambda_{\textsc{n}}$ regime. Of course a larger effective width is still possible if there are 
several nearly degenerate scalar quirk states, which, as briefly mentioned earlier, 
can arise if $\chi$ transforms nontrivially under $\mathrm{SU}(2)_{\textsc{l}}$.

\section*{Other signatures}

While the two photon decay channel of the bound state should be the most
important signature, the dominant decay is expected to be via $\Pi \rightarrow
gg$ and $\Pi \rightarrow \mathcal{H} \mathcal{H}$. The former process is
expected to lead to dijet production while the latter will be an invisible
decay. The dijet cross section is easily estimated:
\begin{align}
  \begin{split} 
  \sigma (p p \rightarrow jj) &\approx
  \begin{cases}
    2.6 N \times \text{Br}(\Pi \rightarrow gg) \text{ pb} & \text{at 13 TeV} \\
    0.5 N \times \text{Br}(\Pi \rightarrow gg) \text{ pb} & \text{at \phantom{1}8 TeV}
  \end{cases}.
  \end{split}
\end{align}
The limit from 8 TeV data is $\sigma(pp \rightarrow jj) \lesssim 2.5$
pb~\cite{Aad:2014aqa, Khachatryan:2015dcf}. If gluons dominate the $\Pi$ decays
(i.e. $\text{Br}(\Pi \rightarrow gg) \approx 1$) then this experimental limit is
satisfied for $N \leq 5$. For sufficiently large $\alpha_{\textsc{n}}$ the
invisible decay can be enhanced, thereby reducing $\text{Br}(\Pi \rightarrow
gg)$. In this circumstance the bound on $N$ from dijet searches would weaken.

The invisible decays $\Pi \rightarrow \mathcal{H} \mathcal{H}$ are not expected
to lead to an observable signal at leading order for much of the parameter space
of interest\footnote{Scalar quirk loops can mediate hueball decays into gluons
  and other \textsc{sm} bosons~\cite{hall, Juknevich:2009ji, Juknevich:2009gg}.
  The decay rate is uncertain, depending on the non-perturbative hueball
  dynamics. However, if the hueballs are able to decay within the detector then
  they can lead to observable signatures including displaced vertices. This
  represents another possible collider signature of the model.}. However, the
bremsstrahlung of a hard gluon from the initial state: $pp \rightarrow \Pi g
\rightarrow \mathcal{H}\mathcal{H} g$ can lead to a jet plus missing transverse
energy signature. Current data are not expected to give stringent limits from
such decay channels, however this signature could become important when a larger
data sample is collected. Note though that the rate will become negligible in
the limit that $\alpha_{\textsc{n}}$ becomes small. Also, in the small
$\Lambda_{\textsc{n}}$ regime, where the bound state is formed from $\chi
\bar{q}$ and $\chi^\dagger q$, the two-body decay $\Pi \rightarrow g \gamma$
($\text{jet} + \text{photon}$) will also arise as in this case the scalar quirk
pair is not necessarily in the color singlet configuration. The decay rate at
leading order is substantial:
\begin{equation}
\frac{\Gamma(\Pi \rightarrow j\gamma)}{\Gamma(\Pi \rightarrow \gamma \gamma)} = \frac{8\alpha_s}{3\alpha Q^2}\ .
\end{equation}
Nevertheless, we estimate that this is still consistent with current data \cite{photon}, but would be expected to become important
when a larger data sample is collected.

Another important signature of the model will be the $pp \rightarrow \Pi \rightarrow Z\gamma$
and $pp \rightarrow \Pi \rightarrow ZZ$ processes. The rates of these decays, relative to $\Pi
\rightarrow \gamma \gamma$, are estimated to be:
\begin{align}
  \begin{split}
  \frac{\Gamma (\Pi \rightarrow Z \gamma) }{\Gamma (\Pi \rightarrow \gamma \gamma)} &= 2 \tan^2 \theta_{\textsc{w}},\\
  \frac{\Gamma (\Pi \rightarrow Z Z) }{\Gamma (\Pi \rightarrow \gamma \gamma)} &= \tan^4 \theta_{\textsc{w}}.
  \end{split}
\end{align}
If $\chi$ transforms nontrivially under $SU(2)_{\textsc{l}}$ then
deviations from these predicted rates arize along with the tree-level
decay $\Pi \rightarrow W^+ W^-$.


\section*{Conclusions}

We have considered a charged scalar particle \(\chi\) of mass
around 375 GeV charged under both $\mathrm{SU}(3)_{\textsc{c}}$ and a new
confining gauge interaction (assigned to be $\mathrm{SU}(N)$ for definiteness). These interactions confine
$\chi^\dagger\chi$ into non-relativistic bound states whose decays into photons
can explain the 750 GeV diphoton excess observed at the LHC. Taking the new confining group to be
$\mathrm{SU}(2)$, we found that the diphoton excess required $\chi$ to have
electric charge approximately $Q \sim [\frac{1}{2}, 1]$. An important feature of our
model is that the exotic particle $\chi$ has a mass much greater than the
$\mathrm{SU}(N)$-confinement scale $\Lambda_{\textsc{n}}$. In the absence of
light $\mathrm{SU}(N)$-charged matter fields this makes the dynamics of this new
interaction qualitatively different to that of \textsc{qcd}: pair production of
the scalars and the subsequent formation of the bound state dominates over
direct bound state resonance production (at least in the perturbative regime
where $\Lambda_{\textsc{n}} \lesssim \Lambda_{\textsc{qcd}}$). Since $\chi$ is a
Lorentz scalar, decays of $\chi^\dagger \chi$ bound states to lepton pairs are
naturally suppressed, and thus constraints from dilepton searches at the
\textsc{lhc} can be ameliorated. This explanation is quite weakly constrained by
current searches and data from the forthcoming run at the \textsc{lhc} will be
able to probe our scenario more fully. In particular,
dijet, mono-jet, di-Higgs and $\text{jet} + \text{photon}$
searches may be the most promising discovery channels.


\section*{Acknowledgements}

This work was supported by the Australian Research Council. Feynman diagrams
were generated using the Ti\textit{k}Z-Feynman package for \LaTeX~\cite{Ellis:2016jkw}.

\end{document}